COMO: A Pipeline for Multi-Omics Data Integration in Metabolic Modeling and Drug Discovery

Brandt Bessell[1$], Josh Loecker[1$], Zhongyuan Zhao[1], Sara Sadat Aghamiri[1], Sabyasachi Mohanty[1], Rada Amin[1], Tomáš Helikar[1,*], Bhanwar Lal Puniya[1,*]

**Affiliations:**
[1]Department of Biochemistry, University of Nebraska-Lincoln, NE

$ These authors contributed equally.
***To whom correspondence should be addressed:**

**Tomáš Helikar**, Ph.D., Department of Biochemistry, University of Nebraska-Lincoln, USA, Email:

thelikar2@unl.edu

**Bhanwar Lal Puniya,** Ph.D., Department of Biochemistry, University of Nebraska-Lincoln, USA, Email: bpuniya2@unl.edu

# Abstract

Identifying potential drug targets using metabolic modeling requires integrating multiple modeling methods and heterogenous biological datasets, which can be challenging without sophisticated tools. We developed COMO, a user-friendly pipeline that integrates multi-omics data processing, context-specific metabolic model development, simulations, drug databases, and disease data to aid drug discovery. COMO can be installed as a Docker image and includes intuitive instructions within a Jupyter Lab environment. It provides a comprehensive solution for multi-omics integration of bulk and single-cell RNA-seq, microarrays, and proteomics to develop context-specific metabolic models. Using public databases, open-source solutions for model construction, and a streamlined approach for predicting repurposable drugs, COMO empowers researchers to investigate low-cost alternatives and novel disease treatments. As a case study, we used the pipeline to construct metabolic models of B cells, which simulate and analyze them to predict 25 and 23 metabolic drug targets for rheumatoid arthritis and systemic lupus erythematosus, respectively. COMO can be used to construct models for any cell or tissue type and identify drugs for any human disease. The pipeline has the potential to improve the health of the global community cost-effectively by providing high-confidence targets to pursue in preclinical and clinical studies.

**Availability and Implementation:** The source code of the COMO pipeline is available at: https://github.com/HelikarLab/COMO

The Docker image can be pulled at:
https://github.com/HelikarLab/COMO/pkgs/container/como
Contact: bpuniya2@unl.edu and thelikar2@unl.edu

## Introduction

Human diseases often lead to changes in cellular metabolism, making metabolic pathways potential targets for treatment (DeBerardinis and Thompson, 2012). Genome-scale metabolic network models (GSMNs) are powerful computational tools that can be used to investigate the effects of drugs on a cell's phenotype and generate testable hypotheses (Jerby and Ruppin, 2012); (Uhlen *et al.*, 2017). In recent years, sophisticated tools such as the COBRA Toolbox have facilitated the construction and analysis of metabolic models (Heirendt *et al.*, 2019; Ebrahim *et al.*, 2013). However, constructing tissue- and cell-type-specific GSMNs and performing systems-level studies often require various tools based on different programming languages, APIs, data resources, and data formats. For example, R/Bioconductor is well-suited for analyzing high-throughput datasets, while Matlab or Python are often used for metabolic modeling and analysis. In addition, model integration with external databases such as DrugBank (Wishart *et al.*, 2018) to run simulations under different conditions often requires additional manual, tedious, and error-prone steps. The complexity and diversity of tools and data resources handled with different programming languages with manual steps could pose significant challenges for researchers, potentially leading to errors, inefficiencies, and delays in using metabolic models for therapeutic interventions and other applications. Furthermore, as the amount of disease-specific omics data has increased in recent years, there has been a growing need for efficient bioinformatics tools to interpret and connect this data. (Dash *et al.*, 2019).

To address these challenges, we developed COMO (COnstraint-based Metabolic Optimization), a pipeline that integrates data sources and analysis tools in a single, easy-to-use platform for researchers to harness the ever-growing volume of publicly available omics data. The pipeline allows users to analyze and integrate multi-omics data to build GSMNs for various tissue and cell types in different biological contexts. It combines transcriptomics and proteomics data from user-provided or publicly available databases using standardized meta-analysis techniques to synthesize constraint-based metabolic models and leverages existing knowledge from drug databases to identify potential therapeutic targets for user-specified diseases. COMO allows rapid, customizable context-specific constraint-based model generation, disease analysis, and drug perturbation analysis. It comes packaged with popular standard tools such as Memote (Lieven *et al.*, 2020), Escher (King *et al.*, 2015), and UMAP (McInnes *et al.*, 2018) for exploratory analysis and clustering.

In this study, we showed an application of the novel COMO pipeline to construct metabolic models of B cells and use them for drug target identification against autoimmune diseases: rheumatoid arthritis (RA) and systemic lupus erythematosus (SLE). Naive B cell metabolism has been shown to play a role in the onset of autoimmune diseases. For example, B cells from SLE patients have altered metabolism because of increased activity of the mTOR pathway (Torigoe *et al.*, 2017) and B cells have already been explored as therapeutic targets against SLE (Atisha-Fregoso *et al.*, 2021). Characterizing the naive B cell metabolism and understanding its role in autoimmune diseases could lead to new therapeutic strategies to restore normal immune function, reduce B cell activity, and prevent the production of harmful antibodies.

# Methods

## Pipeline design

COMO consists of four major steps for identifying drug targets (Figure 1).

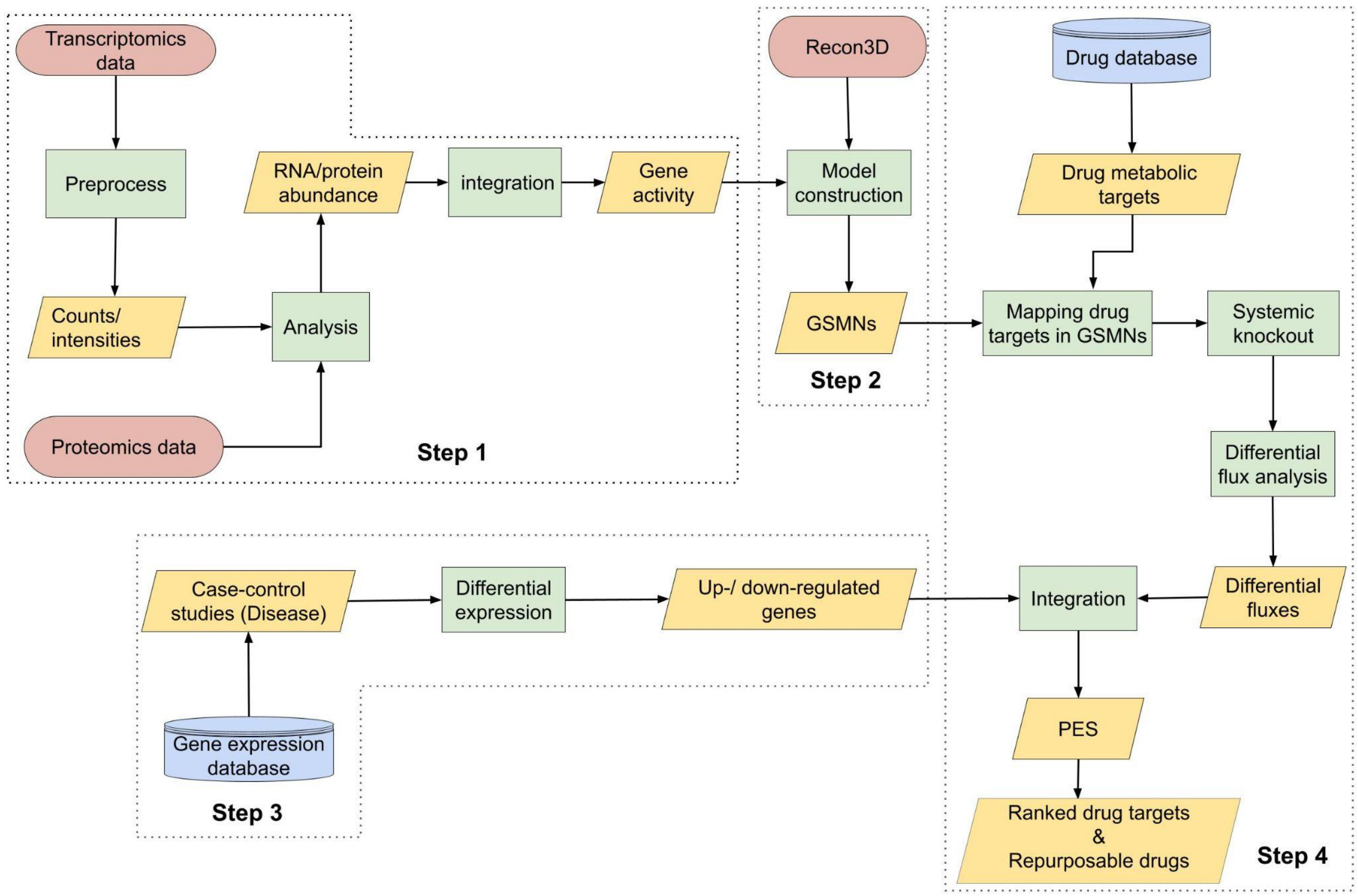

**Figure 1:** Flow chart of the COMO pipeline. GSMN refers to genome-scale metabolic networks.

The first step involves data analysis and consists of three sub-steps: (i) if bulk RNA-seq data is used for constructing the models, preprocessing the data and preparing the configuration files; (ii) analysis of any combinations of bulk-RNA-seq, single-cell RNA-seq, microarrays, and proteomics data, and generation of a list of active genes for each data type; (iii) checking for consensus among user-provided data types according to the desired level of rigor and merging them into a single set of active genes.

The second step uses the list of active genes generated in the third step and a reference model, for example, Recon3D (Brunk *et al.*, 2018), provided in the pipeline to build cell-type-specific models. Users can manually refine these models and provide them in Step 4. Users can also provide their version of the reference model and input datasets, but these must have the same gene annotations (e.g., Entrez IDs or HGNC symbols) if both are user-provided.

The third step analyzes disease-specific data from a user-provided case-control transcriptomics study using bulk RNA-seq or microarrays. COMO uses a platform-based standard pipeline to preprocess, normalize, and identify differentially expressed genes by comparing control groups with patient groups. Ideally, the disease datasets come from the same cell types as the GSMN models, but if data for the specific cell types are unavailable, it advises mixed cell populations data such as PBMCs for T cell lymphocytes.

The fourth step performs the drug perturbation simulation for drug discovery and consists of several substeps: (i) mapping of drug targets obtained from the repurposing tool of the ConnectivityMap database (Corsello *et al.*, 2017) to metabolic genes in the model; (ii) systematic knockouts of each mapped gene; (iii) comparison of flux profiles from perturbed and control models to identify differential fluxes; (iv) comparison of differentially regulated fluxes with differentially expressed genes identified in Step 3 to determine the number of genes up- and down-regulated in a disease whose fluxes are in the opposite direction after perturbation with the mapped drug. (v) computation of the Perturbation Effect Score (PES) (Puniya et al., 2021) for each drug target; the pipeline output includes PES-based ranks of drug targets and mapped repurposable drugs for the studied disease.

## File Preparation and omics data analysis

### RNA-seq Analysis

For RNA-seq analysis, COMO requires a CSV file containing a matrix of gene counts from aligned FastQ reads for each context to synthesize into a context-specific constraint-based metabolic model. This matrix should have column names corresponding to genes such as ENSEMBL, HGNC Symbol, or Entrez ID. Row names should correspond to sample IDs in a specific format (Supplementary Methods). Users can directly use counts matrix files from GEO (Gene Expression Omnibus) (Barrett *et al.*, 2013) and from the supplementary material of publications. However, we recommended using the same alignment and gene quantification pipeline for each sample using the same parameters to maintain consistency and reduce technical variation. A peripheral Snakemake-based pipeline for HPC permits to align, quantify, and run a suite of quality control metrics on FastQ reads from bulk RNA-seq data. It integrates multi-QC, fastq_screen, Trim_galore, picard, and the STAR aligner. We designed the pipeline to interface between GEO and COMO directly. It accepts a CSV file with SRR codes, the sample ID, strandedness, and library preparation technique and outputs a directory formatted with all the files necessary for COMO to analyze and synthesize transcriptomics data.

Users must place all the properly formatted files in designated user data directories to provide input to the COMO pipeline. The outputs of the pipeline will be saved in designated directories. A preprocessing step reads gene counts files and other necessary metrics from each sample, generates and writes a combined counts matrix for each context, creates a file storing important gene-specific information such as gene start points, stop points, and synonyms fetched from BioDBnet (Mudunuri *et al.*, 2009), and creates a configuration file containing the sample ID, batch group, mean fragment length calculated from the output of RNA_fragment_size.py from ReSeQC (Wang *et al.*, 2012), strand specificity, layout, and library preparation technique. This configuration file is used during downstream analysis to get specific values necessary for analysis and to see which samples to include. This makes it easy for the user to exclude samples or batches from the analysis without deleting their associated files from COMO.

COMO can integrate multiple types of RNA-seq data for context-specific GSMN generation. Initial statistical analysis is handled separately for total bulk RNA, polyA enriched bulk RNA, and single cell analysis. For each type of RNA-seq data, gene counts for each sample are normalized, filtered, binarized, and combined using one of 3 user-selected modes: 'zFPKM', 'TPM-quantile', or 'CPM-cutoff' (Supplementary Methods). The output of this step is a CSV file containing all genes in all libraries with binary values for each gene, 0 for inactive and 1 for active. If using zFPKM, COMO will also output a file with combined weighted Z-score transformed values representing zFPKM transformed expression across all replicates in all batches (Supplementary Methods).

### Microarray Analysis

COMO can use microarray datasets from various platforms, including Affymetrix, Agilent, and Illumina. The microarray data is abundant in public repositories and can be used when other advanced data, such as RNA-seq and proteomics, are unavailable for the cell or tissue types of interest. The pipeline uses GEOparser, affy, and limma with a custom-built program for data merging. The user provides a configuration file containing the GEO database's accession numbers of series, samples, and platforms (i.e., GSE, GSM, GPL) and the name of the platform used (i.e., Affymetrix and Agilent) grouped by cell types. The pipeline will automatically download datasets from the GEO database, normalize and analyze them. Data across multiple samples are combined based on binarized expression values and user-defined ratios.

### Proteomics Analysis

Protein abundance data can be used on its own to build context-specific GSMNs, or integrated with transcriptomics data to enhance the model. Because of post-transcriptional, translational, and post-translational regulations, protein abundance is closer to the functional metabolic state of the phenotype than transcriptomics expression data. For transcriptomics and proteomics analysis, technical factors like signal amplification, noise, and degradation are limiting factors when attempting to infer a contextual metabolic state from expression data (Buccitelli and Selbach, 2020).

Proteomics data should be imported into COMO as a CSV or Excel file where the row names are Gene Symbols (standard to MaxQuant machines) or other recognizable formats by BioDBnet, and column names are sample IDs. Each sample should contain protein abundance. A user-specified quantile (25% by default) categorizes the gene(s) corresponding to the proteins with measured abundances into binarized active or inactive states. Similarly to the RNA-seq analysis, binarized activity states can be combined across samples using user-specified ratios.

## Combining Activity of Different ‘Omics’ Data Sources

COMO merges activity states for any combination of available ‘omics’ data sources using binarized data. Users define a minimum activity requirement, which indicates the minimum number of provided sources in which a gene should be active for the gene to be active in the model. Users can thus build a larger, more inclusive, but less strict model that can use any gene active in any provided data source. Users can also build a smaller, stricter, more validated model, potentially missing essential reactions that must be filled in during reconstruction. Users can use high-confidence genes to override the influence of other data sources in the stricter models to ensure they are not excluded from the model. Importantly, users can reduce the minimum activity requirement for each data source with no data associated with that gene to prevent data sources with less measured genes (such as proteomics) from unjustly barring a gene from being active in the final model.

If using weighted combined zFPKM transformed RNA-seq data, protein abundance data can be integrated using the weighted Z-score transform technique but using user-specified weights. Users can thus assign a higher weight to protein abundance data which should be closer to the actual phenotypic metabolic state (Buccitelli and Selbach, 2020).

## Context-Specific Model Extraction

Context-specific models can be seeded from a general genome-scale model, such as Recon3D for human tissue models (Brunk *et al.*, 2018) or iMM1865 for mouse tissue models (Khodaee *et al.*, 2020). These models and similar genome-scale models of general metabolism are readily available for download and can be provided to COMO as SBML files, MAT files created with the COBRA Toolbox in Matlab, and JSON files created by COBRApy. Users should specify a solver. COMO currently supports GLPK (GNU Linear Programming Kit) (Oki, 2012) and GUROBI (Gurobi Optimization, 2022). The solver will be used to subset and solve a context-specific model from the reference global model using a reconstruction algorithm implemented with Troppo (Ferreira *et al.*, 2020).

Users can provide ‘boundary conditions’, ‘force reactions’, and ‘exclude reactions’ files. The boundary conditions file should be a list of exchange reactions, sink reactions, or demand reactions with upper and lower boundaries that can be utilized as inputs for the model, though this will not guarantee that these reactions will be used. If users provide no boundary reactions, the default upper and lower bounds for sinks, exchanges, and demand reactions in the reference model are used, meaning any boundary reaction in the reference model can be used to fill in the reactions mapped from transcriptomics and proteomics data. We recommend that users perform a literature review to define reasonable boundaries for their models to meet a higher degree of physiological relevance. The optional force reactions files can contain a single column of reactions that should be used in context-specific reconstruction, regardless of expression data. We recommend that users populate this file with reactions known to take place in the context based on the literature review. The optional exclude reactions file can contain reactions that should be excluded from the model regardless of expression data.

Metabolic objectives can also be specified, essential for objective-based reconstruction algorithms like GIMME. For non-objective-based reconstruction algorithms like FASTCORE, the objective will be treated as a forced reaction. The gene activity data required for reconstruction depends on the reconstruction algorithm used. Expression distributions are required for iMAT and tINIT reconstructions, whereas binarized gene activity should be used for FASTCORE and GIMME. The selection of a reconstruction algorithm should be based on the goals of the context-specific model and the available data since no reconstruction algorithm has been found to universally yield the most biologically accurate model. (Opdam *et al.*, 2017).

## Identifying Disease-Related Genes

Disease-related genes can be determined using differential gene expression analysis (currently with RNA-seq or microarray data). For RNA-seq data, COMO requires a configuration file specifying sample names annotation for patient and control samples. The sample names should correspond to column names in a gene counts matrix which can be downloaded from GEO or processed from the output of the Snakemake alignment pipeline discussed previously. Differential gene expression is performed using the edgeR (Robinson *et al.*, 2010) or DESeq2 (Love *et al.*, 2014) Bioconductor packages. For microarrays, COMO requires a user-supplied configuration file containing GEO accessions (i.e., GSE, GSM, and GPL) and the name of the platform (e.g., Affymetrix, Agilent) with the patient and control sample annotation. The data are downloaded from GEO and analyzed using the affy (Gautier *et al.*, 2004) and limma (Ritchie *et al.*, 2015) Bioconductor packages based on the configuration file. COMO uses the platform-based (such as Affymetrix and Agilent) standard analysis workflow using Bioconductor to preprocess, normalize, and identify differentially expressed genes by comparing control and patient groups. This step outputs genes that are up-regulated and down-regulated.

## Identifying Drug Targets and Repurposable Drugs

Drug targets obtained from the ConnectivityMap database (Subramanian *et al.*, 2017) are mapped to metabolic genes in the model, where systematic knockouts are performed on each gene. Differential fluxes are identified by comparing flux profiles of perturbed and control models. Differentially regulated fluxes are compared with differentially expressed genes identified in the previous step. This comparison determines the number of up- and down-regulated genes in a disease reversed by each mapped drug. The Perturbation Effect Score (PES, (Puniya *et al.*, 2021)) is calculated for each drug target. PES combines two values: the number of down-regulated fluxes controlled by up-regulated genes minus up-regulated fluxes controlled by up-regulated genes and the number of up-regulated fluxes controlled by down-regulated genes minus down-regulated fluxes controlled by down-regulated genes. These two values are normalized with the total number of fluxes and then combined. The output of this step includes PES-based ranks of drug targets and mapped repurposable drugs for the disease of interest.

All the available features in the COMO pipeline are given in Table 1.

**Table 1: Available features in the COMO pipeline**

| **Features in COMO** | |
|---|---|
| **Data integration** | **Model seeding** |
| Microarray | iMAT |
| Bulk RNA-seq | FastCORE |
| Single-cell RNA-seq | GIMME |
| Protein abundance | |
| **Peripheral analysis (Docker container)** | **Peripheral Analysis (HPC Snakemake pipeline)** |
| Differential Gene Expression | Fastq_screen integration |
| Drug target analysis (inhibition) | FastQC integration |
| Sample clustering | Trim_galore integration |
| Memote integration | Picard RNA-seqMetrics integration |
| Escher mapping integration | MultiQC integration |
| | STAR alignment integration |

## B cell data preprocessing and analysis

We applied the COMO pipeline to drug discovery in B cells. We collected RNA-seq and protein abundance datasets published in GEO (Barrett *et al.*, 2013). We processed the RNA-seq datasets using our accompanying alignment and quality control pipeline on the University of Nebraska-Lincoln's high-performance computing cluster, 'Rhino'. We selected the RNA-seq datasets based on criteria defined in Supplementary Methods. Finally, we selected three RNA-seq datasets with ten samples for COMO to analyze and merge to construct context-specific models (Table 2). We also used published quantitative proteomics data (protein abundance) generated based on high-resolution mass spectrometry. We used four replicates for naive B cells for model construction (Table 2).

**Table 2: CBM model bulk RNA-seq and proteomics datasets**

| Data type | GSE accession number or Reference | Cell Type | Number of QC Passing Replicates |
|---|---|---|---|
| Bulk RNA-seq | GSE118165 | Naive B | 4 |
| | GSE107011 | Naive B | 4 |
| | GSE92387 | Naive B | 2 |
| MS-based proteomics data | Rieckmann et al., 2017 | Naive B | 4 |

We imported RNA-seq read counts and protein copy number data to the COMO container and processed them using the following procedure, which can be replicated using the provided Juypter notebook file. The function 'rnaseq_preprocess.py' was used to generate count matrix files. The automatically generated configuration .xlsx files were modified to exclude RNA-seq datasets, which failed quality control checks shown in Supplementary Table 1. The parameters used in B cell data analysis and integration are shown in Supplementary Table 2. The RNA-seq counts and protein copy number values were transformed to Z scores. Weights for RNA-seq-based Z-transformed scores and proteomics-based Z-transform scores were 6 and 10, respectively, to follow the 0.6 correlation coefficient found between transcription and protein expression by (Buccitelli and Selbach, 2020).

# Results

## Constructing metabolic models

We used the combined z-transformed values based on transcriptomics and proteomics data of naive B cells (Figure 2A) to seed a model using the iMAT and GIMME algorithms in conjunction with the GUROBI solver. We used biomass maintenance for the objective function of naive B cells since they do not proliferate (Moscardó García *et al.*, 2021). We defined boundary exchange reactions based on literature about B cells' nutrient preferences. We determined forced reactions from the literature reporting experimentally validated reactions active in B cells. The boundary and forced reactions are provided in Supplementary Table 3. Using combined RNA-seq and proteomics datasets, we generated many models at different data cutoffs ranging from combined Z-transformed value -6 to 4. To assess the functionality of the default zFPKM cutoff of -3 to -2 proposed by (Hart *et al.*, 2013), we generated calibration models using GIMME and iMAT methods. For GIMME models, we started from the lowest functional cutoff of >-4 created from binarized data. For iMAT calibration models, we used >-5 as a high threshold in combined Z-transform scores. A lower threshold of <-5 was also used to prefer using genes that did not show near zero expression. This will fulfill a minimum flux requirement through reactions associated with genes expressed above the higher threshold. The GIMME algorithm results in models with infeasible and dead-end reactions, which must be made fixed using an algorithm such as FASTCC (Vlassis *et al.*, 2014). The COBRApy version of FASTCC can be used in COMO. We used sanity checks (Heirendt *et al.*, 2019) to determine baseline characteristics for each calibration model, including metabolic tasks and reactions contributing to biomass normalized by total feasible reactions.

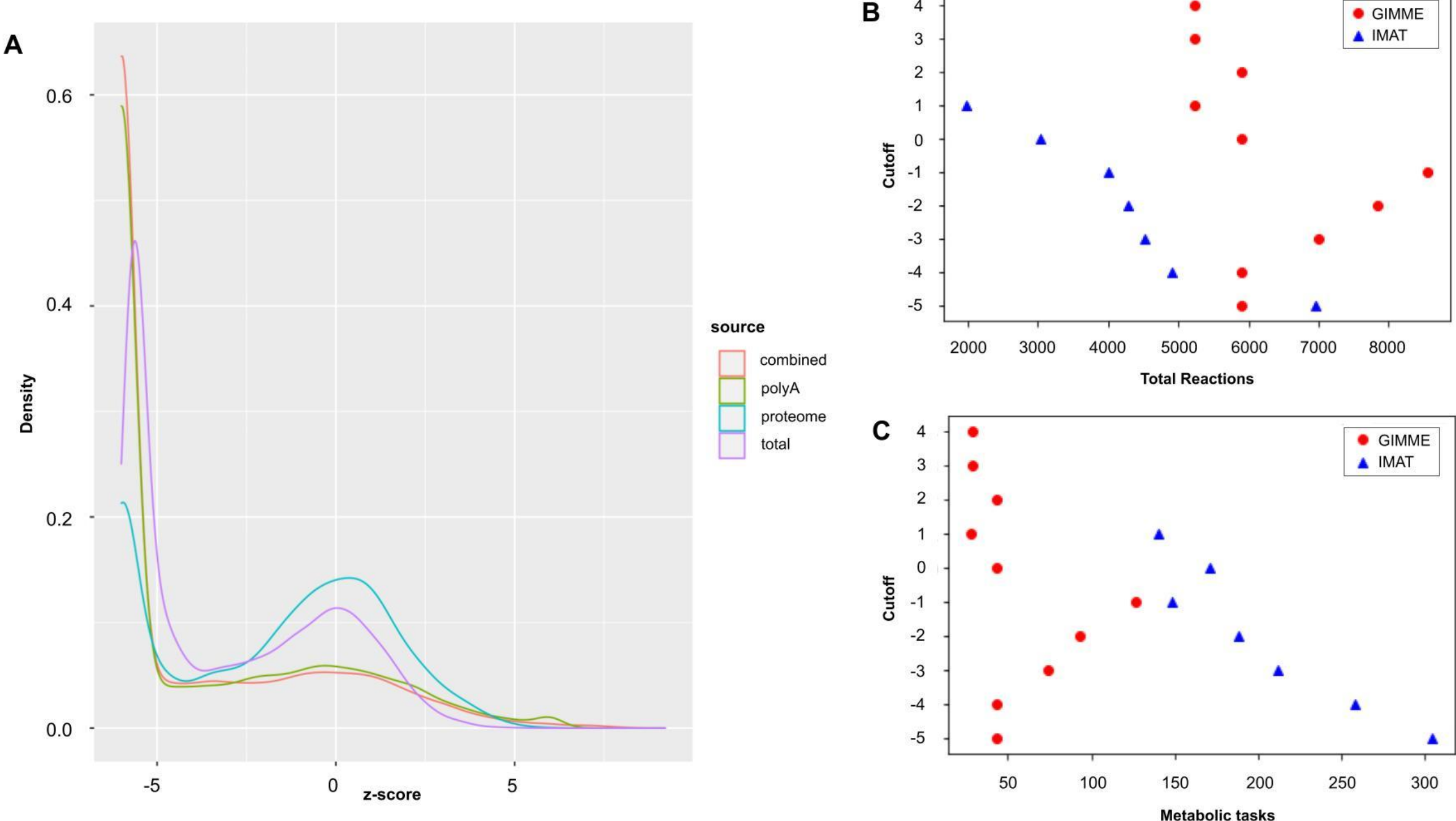

**Figure 2:** Combined Z-transform scores and their effect on model building and performance. (A) Combined Z-transformed scores for different B cell datasets. (B) Effect of Z-score cutoffs on the total number of reactions in the model constructed using GIMME and iMAT algorithms. (C) Number of metabolic tasks completed by GIMME and iMAT-based models constructed with different Z-score cutoffs.

We used results from physiological tests and sanity checks to evaluate the models and advance them for further analysis. To choose the best multi-omics data cutoff, we investigated (i) the number of feasible reactions (Figure 2B) and the number of metabolic tasks passed (Figure 2C) and (ii) the number of reactions in the final model that contributed to maximizing the objective functions (Figure 2C and Supplementary Table 4). The ratio of metabolic tasks versus feasible reactions can help us identify the contextualized model with better retention of global metabolic functions while ensuring its reduction compared to the reference model (as the previously published CD4+ T cell models). A higher number of reactions contributing to the objective functions can indicate greater connectivity of reactions within the network. A Z-transformed value of -3 generated the best iMAT model with 4528 feasible reactions and 212 metabolic tasks passed. A -3 cutoff has also generated the best GIMME model with 6999 reactions and 74 metabolic tasks passed. Based on the sanity checks of selected iMAT and GIMME models, individually investigated, iMAT models showed better performance regarding the number of completed metabolic tasks and ATP production from different carbon sources (Supplementary Data 1). The parameters used for constructing models are provided in Supplementary Table 2.

## Metabolic model of naive B cell

The selected iMAT-based naive B cell model passed all the sanity checks and completed 212 metabolic tests out of 460 human metabolic tasks (Supplementary Table 4). This model comprises 4528 reactions and 1463 genes (Supplementary Data 2). 4064 reactions were internal enzyme-catalyzed and transport reactions distributed across 93 metabolic pathways in different compartments (extracellular, cytosol, mitochondria, endoplasmic reticulum, Golgi apparatus, lysosome, nucleus, and peroxisome). Most reactions were involved in extracellular transport, followed by fatty acid oxidation, mitochondrial transport, peptide metabolism, and fatty acid synthesis (Supplementary Figure 1).

We validated the naive B cell model by comparing it with published literature on B cell metabolism. We used Flux Balance Analysis (FBA) to identify active pathways and Minimization of Metabolic Adjustment (MoMA) to simulate knockouts. The results presented in Table 3 highlight the agreement between our model and the established biological scenarios in the literature. Specifically, our FBA analysis revealed that several major pathways agreed with the literature, including glycolysis, lactate production, the TCA cycle, and glucose transport (see the illustrative flux map of these pathways in Figure 3). Additionally, in agreement with the literature, our model did not show any impact of glucose removal from the exchange media on growth (Figure 4A). The literature has shown that inhibition of HIF1 induces apoptosis and decreases mature B cells (Boothby *et al.*, 2022). HIF-1 increases the expression of glycolytic enzymes (Taylor and Scholz, 2022). To simulate HIF-1’s effect on B cells, we knocked out its target genes. GAPD knockout reduced B cell growth to zero, and other enzymes also affected growth (Figure 4B). It is worth noting that we identified the active pathways and gene essentiality based on the literature and investigated the activity of the corresponding models with FBA. This approach allowed us to validate our models against known biological scenarios of naive B cell metabolic pathways.

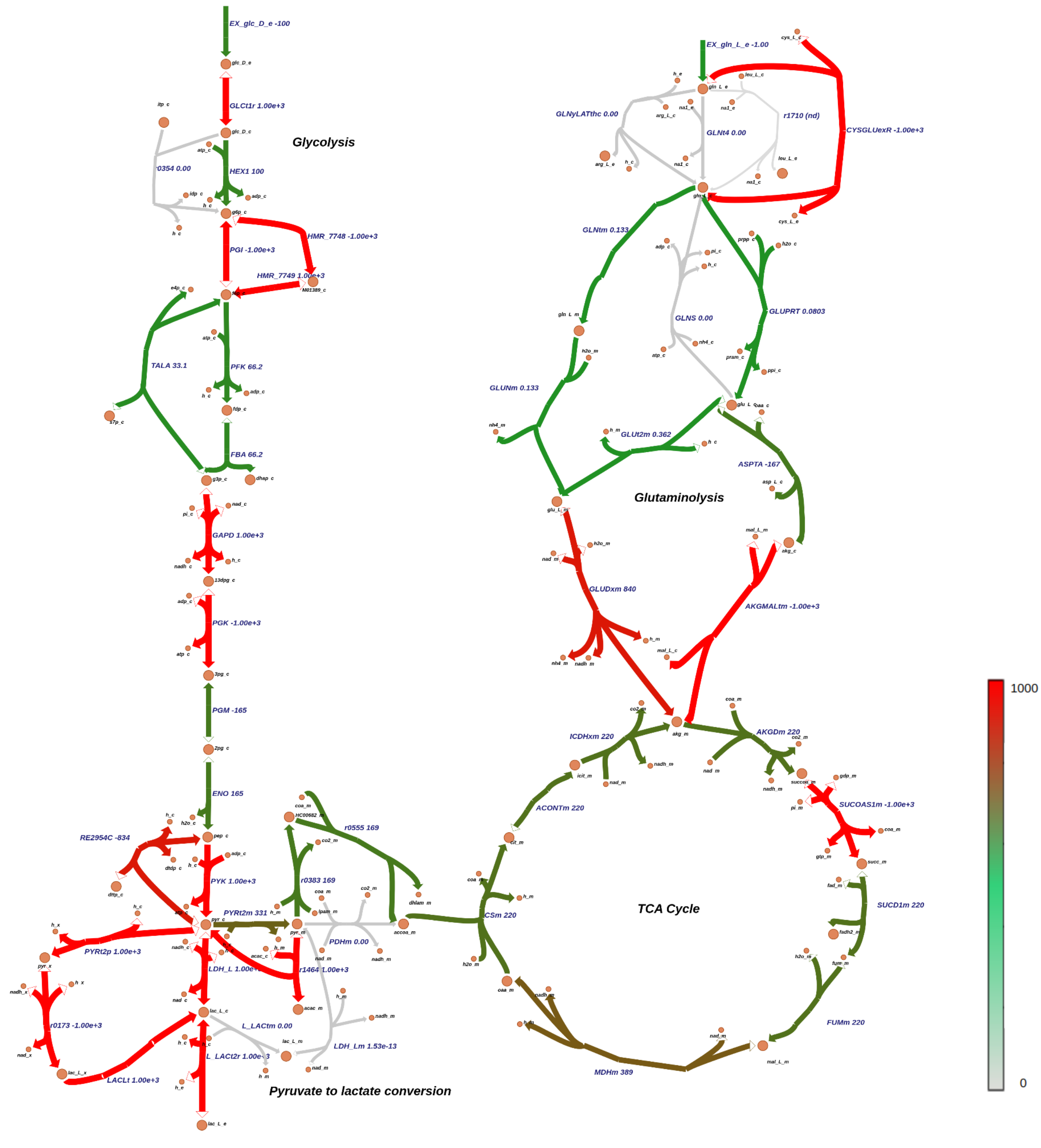

**Figure 3: Flux distribution through glycolysis, TCA cycle, and glutaminolysis pathways.**

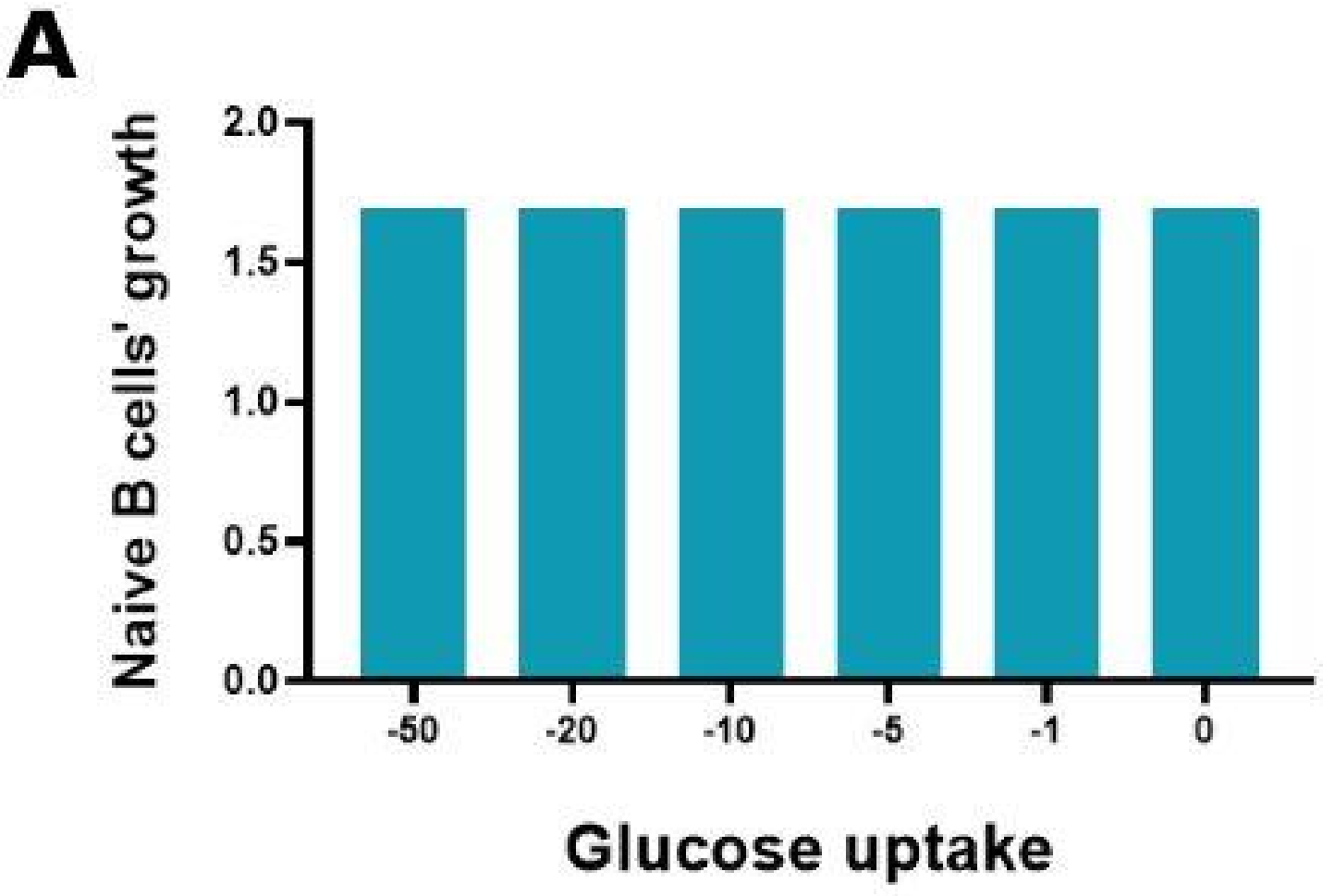

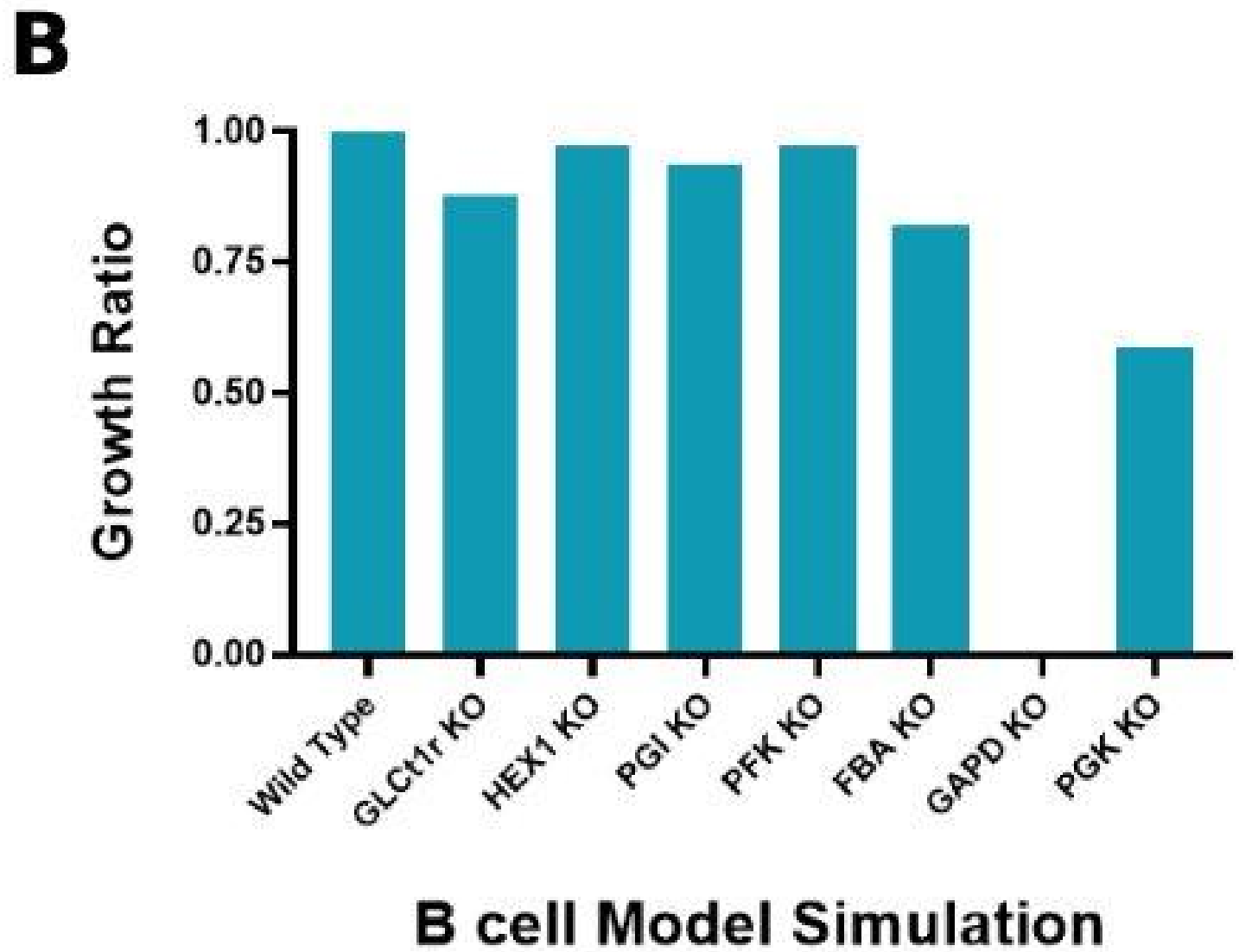

**Figure 4: Validation of B cell model.** (A) Limiting glucose in media has no impact on naive B cell growth. (B) Impact of glycolytic enzymes knockout (growth ratio compared to wild-type). All the tested reactions showed some impact on B cell growth.

**Table 3: Model validation using specific behaviors.**

| # | Observation from literature | *In silico* experiment | Agreement |
|---|---|---|---|
| 1 | Glucose restriction has no impact on naive B cell function. (Waters *et al.*, 2018) | Correlated increased glucose uptake with naive B cells growth | Yes |
| 2 | Glucose distributes through G6P, F16BP, G3P, and 3PG. (Waters *et al.*, 2018) | Checked the presence of the reactions leading to these products | Yes |
| 3 | Naive B cells produce lactate (Waters *et al.*, 2018) | Checked flux through reactions producing lactate | Yes |
| 4 | In naive B cells, citrate, glutamate, glutamine, α-ketoglutarate, succinate, fumarate, malate are active (Waters *et al.*, 2018) | Checked TCA pathway activation | Yes |
| 5 | Untreated or anti-IgM treated B cells showed no significant increase in the ECAR in response to the glucose addition. (Boothby *et al.*, 2022; Akkaya *et al.*, 2018) | Correlated glucose increase with the naive B cells lactate production | Yes |
| 6 | B cell survival requires proper maintenance of glycolytic and oxidative glucose pathways. Inhibition of HIF1 induces apoptosis and decreases mature B cells (Boothby *et al.*, 2022) | Inhibited glycolytic enzymes and check B cell growth | Yes |
| 7 | Naive B cells seem to rely on FA as their main fuel (Wilson and Moore, 2020) | Checked the effect of fatty acids on B cell growth | Yes |

## Disease gene analysis

We collected a publicly available case-control RNA-seq dataset (GSE110999) of naive B cells for two autoimmune diseases, rheumatoid arthritis (RA) and systemic lupus erythematosus (SLE). This dataset included four samples from each disease's patients and healthy controls (Supplementary Table 5). We used COMO's count matrix and configuration files to identify up and down-regulated genes (adjusted P-values<0.05) using EdgeR ('disease_analysis.py', see Supplementary Table 2)

## Drug target analysis

We mapped drug targets in metabolic models using connectivityMap's list of drugs and their targets. A Drug-Target file containing inhibitors and antagonists is included in the pipeline. We simulated knockouts by inhibiting existing drugs' metabolic targets ('knock_out_simulation.py') and compared the simulation results with genes up and down-regulated in diseases, and computed the perturbation effect score (PES, (Puniya *et al.*, 2021)). Finally, we used genes with positive PES for each cell and disease combination as potential targets and their associated drugs as potential candidates for repurposing.
We used the naive B cell metabolic model to identify potential drug targets and repurposing candidates for treating RA and SLE. We integrated differentially expressed genes with model analysis to compute PES and identify targets with drugs already prescribed or studied for autoimmune disorders. A positive PES indicates that reactions associated with differentially expressed genes – when inhibited by drugs – have fluxes in directions opposite to the differential expression. Thus, reactions associated with up-regulated genes have reduced flux, while reactions associated with down-regulated genes have increased flux. We identified 44 genes for RA and 52 for SLE with positive PES. These genes were associated with 123 and 120 drugs for RA and SLE, respectively. Among all genes with positive PES, we found 29 targets common between RA and SLE. A higher PES indicates that the gene's perturbation has a stronger effect on differentially expressed genes. To select genes with high PES, we transformed the PES to Z-scores and removed genes with a z-score lower than 0.5. This resulted in 25 and 23 drug targets for RA and SLE, respectively.

Among the potential targets identified for RA, APRT (coagulation factor inhibitor) was ranked the highest, followed by GLB1, CTSA, ODC1, TBXAS1, KHK, ADK, ALDH5A1, and ACACA (Supplementary Table 6). We identified five genes (TBXAS1 - rank 5, ABCB1 - rank 10, HPRT1 - rank 13, SLC10A1 - rank 33, SOAT1 - rank 40) targeted by the drugs already used to treat RA or are under study for RA. Three targets (TBXAS1, ABCB1, HPRT1) were also in the high-positive PES target list. The drug targets supported by the literature are shown in (Table 4). The new potential drug targets included GLB1 (galactosidase beta 1), ODC1 ( Ornithine Decarboxylase), and ALDH5A1 (Aldehyde dehydrogenase 5 family member A1).

For SLE, GGPS1 (geranylgeranyl diphosphate synthase 1), GSR (glutathione-disulfide reductase), and GLB1 (galactosidase beta 1) were ranked the top three drug targets (Supplementary Table 6). In addition, PPAT (phosphoribosyl pyrophosphate amidotransferase; rank 7) and DHFR (dihydrofolate reductase; rank 23) are the targets of drugs used for SLE (Table 4). The other drug targets included GGPS1 (Geranylgeranyl diphosphate synthase 1), GSR (Glutathione-disulfide reductase), ALDH5A1 (Aldehyde dehydrogenase 5 family member A1).

**Table 4.** Predicted drug targets validated with literature

| Disease | Drug target | Support |
|---|---|---|
| RA | TBXAS1 (Thromboxane A Synthase 1) | Potential target of anti-rheumatic drug Sulfasalazine (Stenson and Lobos, 1983) |
| | ABCB1 (ATP binding cassette subfamily B member 1) | Target of macrolides used as RA treatment option (Wishart *et al.*, 2018; Saviola *et al.*, 2007) |
| | HPRT1 (hypoxanthine phosphoribosyltransferase 1) | Target of mercaptopurine used for RA treatment (Dean and Kane, 2012) |
| SLE | PPAT (Phosphoribosyl pyrophosphate amidotransferase) | Target of the drug Azathioprine used in SLE treatments. (Abu-Shakra and Shoenfeld, 2001; Zhou *et al.*, 2022) |
| | DHFR (Dihydrofolate reductase) | Target of the drug methotrexate, used to treat SLE. (Wishart *et al.*, 2018) |

Next, we investigated the biological processes that are common targets in both diseases. A comparison of drug targets between RA and SLE identified 12 common drug targets: GLB1, CTSA, ALDH5A1, ABCB1, ABCC1, ABCA1, GSR, GCLM, ADH5, GART, PPAT, and ATIC. The gene enrichment analysis (Zhou *et al.*, 2019) showed that the sphingolipid translocation (ABCA1, ABCC1, ABCB1, GLB1, CTSA), and *de novo* IMP biosynthetic process (ATIC, GART, PPAT, GSR, ABCC1, GCLM), were enriched highest in the commonly found genes (Figure 5).

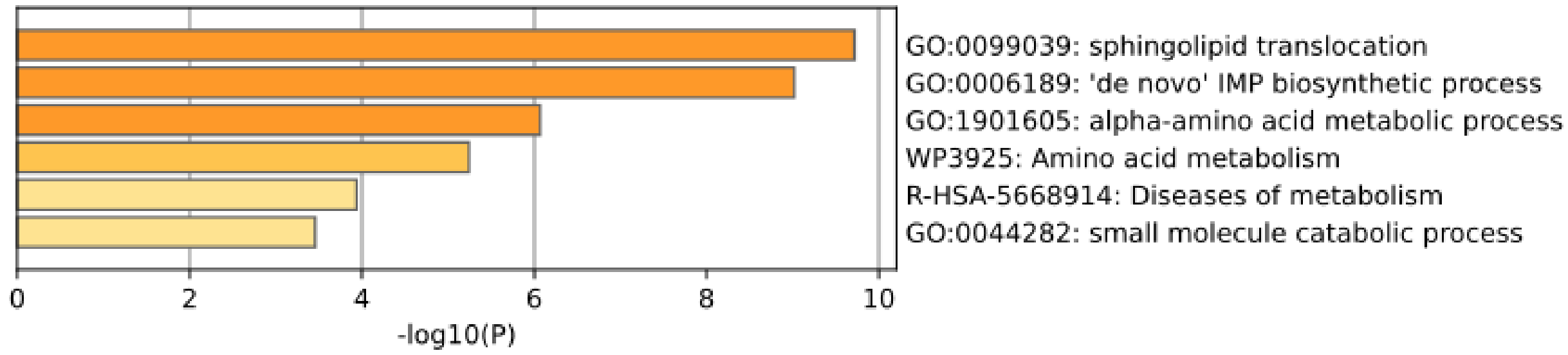

**Figure 5.** Biological processes and pathways enriched in common drug targets between RA and SLE.

# Discussion

Dysregulation of metabolic pathways is often associated with various diseases, making them attractive targets for drug development (Du *et al.*, 2022; Piranavan *et al.*, 2020). For example, metabolic pathways have been found to be disrupted in autoimmune diseases, leading to the production of harmful immune cells or the activation of inflammatory pathways (Piranavan *et al.*, 2020). Computational modeling of metabolism can help simulate the effects of gene or reaction perturbations *in silico* to identify potential drug targets. However, the systematic analysis of such models requires integrating various modeling methods and biological datasets. *In silico,* gene or reaction perturbations can be used to predict how a particular drug may impact a metabolic pathway and the effect on a cell's phenotype. These models can be constructed using various methods and biological datasets, such as transcriptomics and proteomics data. They can be subsequently used to simulate the effects of different drugs on a pathway.

COMO (Constraint-based Optimization of Metabolic Objectives) is a user-friendly pipeline that integrates various modeling methods and biological datasets to process transcriptomics and proteomics data, build context-specific metabolic models, run simulations, and investigate gene inhibition's effect on reactions. It also includes features for identifying repurposable drugs and predicting drug targets. By interfacing with public databases and utilizing open-source solutions for model construction, COMO aims to enable researchers to investigate low-cost alternatives and novel prophylaxis for diseases to improve the health of the global community. We previously used this approach in identifying drug targets in the metabolism of CD4+ T cells (Puniya *et al.*, 2021) defined based on microarrays and proteomics datasets. The COMO pipeline is scalable and robust, which includes single-cell RNA-seq, bulk RNA-seq, and proteomics data, as well as microarrays, can be used to identify active genes in a specific cell or tissue type. By integrating data from single-cell RNA-seq, which provides a high-resolution view of gene expression at the individual cell level, with bulk RNA-seq, which provides a more global view of gene expression in a population of cells, researchers can gain a more comprehensive understanding of the metabolic state of a cell or tissue. Additionally, the integration of multi-omics data, including transcriptomics and proteomics, allows for a more comprehensive understanding of the metabolic state of a cell or tissue and can provide additional insights into the effects of perturbations on the cell's phenotype.

Genome-scale metabolic models (GSMNs) can be used in drug discovery to identify potential therapeutic targets by simulating perturbations in metabolic pathways and analyzing the resulting changes in the cell's phenotype (Turanli *et al.*, 2019; Puniya *et al.*, 2021). GSMNs rely on the idea that a cell's metabolism is constrained by the availability of nutrients and energy and regulatory mechanisms such as enzyme activity and gene expression. Integrating GSMNs with information about known drugs and their effects on cell metabolism makes it possible to identify potential therapeutic targets for treating various diseases. One approach to integrating GSMNs with drug discovery is to simulate the effects of different drugs on cell metabolism and then use the resulting data to identify potential therapeutic targets (Puniya *et al.*, 2021; Larsson *et al.*, 2020). Another approach is to use GSMNs to predict the effects of drugs on specific pathways or processes in the cell and then use this information to design new drugs or repurpose existing drugs to treat specific diseases (Jamialahmadi *et al.*, 2022). One can also check if the predicted targets could be targeted in other immune cells possibly important for the disease. For example, we identified three target genes (CAT, COMT, FDPS) in B cells previously identified as RA drug targets in CD4+ T cells (Puniya *et al.*, 2021). Overall, integrating GSMNs with drug discovery can help identify new therapeutic targets and optimize the development of drugs to treat various diseases.

One advantage of an all-in-one solution for drug discovery in metabolism is the ability to streamline and reduce the time and resources required for data collection and analysis. By integrating various modeling methods and data sources in a single platform, researchers can easily access and analyze a wide range of data, including transcriptomics, proteomics, and drug information, without switching between different tools or platforms. Researchers can thus identify and evaluate potential therapeutic targets and drugs more efficiently, enabling them to advance their research and bring new treatments rapidly. In addition, an all-in-one solution can provide a more cohesive and consistent approach to drug discovery, helping to ensure that the results are accurate and reliable. An all-in-one solution for drug discovery in metabolism can provide significant benefits for researchers and the global community by enabling the development of novel, cost-effective treatments for complex diseases.

COMO presents some limitations that should be considered for drug target identification and repurposing. Firstly, COMO relies on the accuracy and completeness of the available omics data and reference models. If the data are noisy or incomplete, the resulting models and predictions may not be accurate. Secondly, COMO is limited to the scope of the reference models and databases it interfaces with and will not be able to identify targets or repurposable drugs that are not included in these resources.

We used COMO for drug discovery in B cell metabolism against RA and SLE. In RA, TBXAS1 (Thromboxane A Synthase 1), ABCB1 (ATP binding cassette subfamily B member 1), and HPRT1 (hypoxanthine phosphoribosyltransferase 1) were supported by literature evidence. TBXAS1 is a target of Sulfasalazine, an older disease-modifying anti-rheumatic drug (DMARD), currently prescribed as a second or third-line defense against RA. In the serum of RA patients, Thromboxane B2 (product of gene TBXAS1) was significantly higher compared to healthy controls (Wang *et al.*, 2015). ABCB1 is a target of macrolides such as erythromycin used to treat RA. The reason was initially thought to be the control of periodontal bacteria, which may partly be responsible for RA progression (Ogrendik, 2013). However, later studies, such as (Saviola *et al.*, 2007), posit that an anti-inflammatory mechanism is more likely responsible since the discontinuation of clarithromycin results in a rapid rebound of RA symptoms after six months. HPRT1 is the target of azathioprine and mercaptopurine. The latter have been used as effective treatments for RA and SLE. Metabolites of each prodrug enter the cell (Fotoohi *et al.*, 2006), where they are broken down in the purine salvage pathway to form metabolites that inhibit purine synthesis (Karran and Attard, 2008), possibly through non-competitive inhibition of phosphoribosylpyrophosphate amidotransferase (PPAT) (Tay *et al.*, 1969).

In SLE drug targets, PPAT (Phosphoribosyl pyrophosphate amidotransferase ) and DHFR (Dihydrofolate reductase) were supported by the literature. PPAT catalyzes the formation of phosphoribosyl pyrophosphate (PRPP), a key intermediate in the *de novo* purine biosynthesis pathway. PPAT is a target of the drug Azathioprine, an immunosuppressive drug used in SLE treatments. However, this drug has not been studied in B cells. Purine biosynthesis is important for the proliferation and function of immune cells, and targeting this pathway could impact the immune system. DHFR is an enzyme that plays a role in the metabolism of folic acid, and it's an essential enzyme in the metabolic pathway that converts the nutrient folate into its active form. DHFR is known to be a target of the drug methotrexate, used to treat SLE. However, DHFR has not been investigated as a drug target for B cells in systemic lupus erythematosus (SLE). Some studies suggest that folate metabolism could be involved in the pathogenesis of SLE.

For RA, our study predicted 22 new targets, several of them supported by literature experiments demonstrating the anti-inflammatory effect of these genes on various immune cells. For example, GLB1 (galactosidase beta 1) encodes an enzyme that hydrolyzes terminal beta-linked galactose residue from ganglioside substrates and other glycoconjugates. D-Fagomine has been found to control inflammatory processes related to an overactivation of the humoral immune response (Segura and Torras, 2014). ODC1 ( Ornithine Decarboxylase) catalyzes the decarboxylation of ornithine. In type 3 innate lymphoid cells (ILC3s), deficiency of ODC1 reduced IL-22 production and protected mice from colitis (Peng *et al.*, 2022). ALDH5A1 breaks down gamma-aminobutyric acid (GABA). ALDH5A1 inhibition makes more GABA available, which might lead to suppressing the immune response since B cells have been shown to produce GABA  (Zhang *et al.*, 2021).

We predicted 21 potential drug targets for SLE, many of which have been previously studied and linked with SLE. GGPS1 **(**Geranylgeranyl diphosphate synthase 1) is a protein involved in geranylgeranyl diphosphate biosynthesis, a precursor of isoprenoid synthesis. The isoprenoid pathway was found to be up-regulated in SLE patients (Kurup and Kurup, 2003). However, GGPS1 has not been studied as a drug target in B cells. GSR **(**Glutathione-disulfide reductase) catalyzes the reduction of glutathione disulfide (GSSG) into glutathione (GSH). GSH is an important antioxidant that plays a role in protecting cells from oxidative stress. GCLM: Glutamate-cysteine ligase (GCL) is the rate-limiting enzyme in the biosynthesis of the antioxidant glutathione (GSH). GCL comprises two subunits, the catalytic subunit (GCLC) and the modifier subunit (GCLM). GCLM inhibition decreases the GSH, increasing ROS production and affecting T cell differentiation (Lian *et al.*, 2018). ALDH5A1 (Aldehyde dehydrogenase 5 family member A1)  and ABAT (4-aminobutyrate aminotransferase) break down gamma-aminobutyric acid (GABA). ALDH5A1 inhibition makes more GABA available, which might lead to suppressing the immune response since B cells have been shown to produce GABA (Zhang *et al.*, 2021). LTA4H (Leukotriene A4 hydrolase**)** converts leukotriene A4 (LTA4) to leukotriene B4 (LTB4), a potent proinflammatory mediator (Fourie, 2009). LTB4 is involved in the recruitment and activation of immune cells, such as neutrophils and macrophages, and plays a role in inflammation and immune responses. FADS1 (Fatty acid desaturase 1) converts saturated fatty acids to monounsaturated fatty acids. It also regulates the production of certain eicosanoids, which are proinflammatory mediators (Gromovsky *et al.*, 2018) linked to the pathogenesis of various inflammatory and autoimmune diseases, including systemic lupus erythematosus (SLE). ACACB (acetyl-CoA carboxylase beta**)** study suggested the role of ACACB in the pathogenesis of SLE (Harley *et al.*, 2017).

In summary, using metabolic pathways as drug targets can provide new therapies for complex diseases such as autoimmunity and cancers. COMO can aid in identifying potential drug targets by integrating various modeling methods and biological datasets and providing a streamlined approach for predicting repurposable drugs. This can help researchers investigate low-cost alternatives and novel prophylaxis for diseases to improve the global community's health.

**Acknowledgments:** This material is based upon work supported by the Defense Health Agency and U.S. Strategic Command under Contract No. FA4600-12-D-9000 and Contract No. FA4600-18-D-9001, and in part by the National Institutes of Health, grant #R35GM119770, and by the Layman seed grant by University of Nebraska Foundation. Any opinions, findings and conclusions or recommendations expressed in this material are those of the author(s) and do not necessarily reflect the views of the Defense Health Agency, U.S. Strategic Command, or 55th Contracting Squadron. Similarly, the opinions or assertions contained herein are the private views of the authors and are not necessarily those of the Uniformed Services University of the Health Sciences, or the Department of Defense. The mention of specific therapeutic agents does not constitute endorsement by the U.S. Department of Defense, and trade names are used only for the purpose of clarification.